\documentclass[final,floatfix,twocolumn,pre,aps,floats,amsfonts,amssymb,showpacs,amsmath]{revtex4}

\usepackage[final]{graphicx}  

\newcounter{saveeqn}

\begin{document}
\title{Kinematic dynamo simulations of von 
K\'arm\'an flows: application to the VKS experiment}
\author{A.~Pinter}
\email{kontakt@alexander-pinter.de}
\author{B.~Dubrulle}
\author{F.~Daviaud}
\affiliation{Service de Physique de l'Etat Condens\'e,
CEA, IRAMIS, CNRS/URA 2464, CEA-Saclay,
F-91911 Gif-sur-Yvette, France}
\date{\today}

\begin{abstract}
The VKS experiment has evidenced dynamo action in a highly turbulent liquid sodium 
von  K\'arm\'an flow  [R. Monchaux et al., Phys. Rev. Lett. {\bf 98}, 044502 (2007)].
However, the existence and the onset of a dynamo happen to depend on the exact experimental configuration. 
By performing kinematic dynamo simulations on real flows, we study their influence on dynamo action, 
in particular the sense of rotation and the presence of an annulus in the shear layer plane. 
The 3 components of the mean velocity fields are measured in a water prototype for different VKS configurations 
through Stereoscopic Particle Imaging Velocimetry. Experimental data are then processed 
in order to use them in a periodic cylindrical kinematic code. 
Even if the kinematic predicted mode appears to be different from the experimental saturated one, 
the results concerning the existence of a dynamo and the thresholds are in qualitative agreement, 
showing the importance of the flow characteristics.
\end{abstract}

\date{Eur. Phys. J. B 74, 165-176 (2010)}

\maketitle

\section{Introduction}
Dynamo action is an instability by which mechanical energy is transformed into magnetic energy for 
$R_m \ge R_m^c$ where $R_m$ is the magnetic Reynolds number.
Since Larmor's prediction concerning the solar dynamo\cite{Larmor}, a lot of work has 
been devoted to this field, including both theories and numerical simulations.
In the last decade, the interest for dynamo has been renewed by the positive results
obtained by the Riga \cite{G01} and Karlsruhe \cite{SM01} fluid dynamo experiments.
These experiments rely on analytical predictions by Ponomarenko \cite{Pono} and Robert
\cite{GORoberts} and have been designed accordingly.  In particular, the generated turbulent sodium 
flows are organized so that the instantaneous flow remains always close to the mean flow,
allowing the kinematic dynamo predictions of the thresholds and of the neutral modes 
(respectively oscillatory and stationary) to be very accurate.  

A new step was passed very recently in the von K\'arm\'an Sodium (VKS) experiment.
Different dynamos have been observed inside a flow
of liquid sodium at large Reynolds number ($Re\sim 10^7$) \cite{MBB07,BMF07,P3,P5}.
Depending on the the control paramaters, statistically stationary dynamos or dynamical dynamo
regimes including field reversals have been reported. 
Under these conditions, the flow is fully turbulent, with many large and small scale
fluctuations, and the instantaneous flow is very different from the mean flow  \cite{RCD08,CDM09}.
The dynamo instability taking place in the VKS experiment is thus at variance with the classical 
instabilities taking place in laminar flows or in turbulent flows remaining very close to their mean.
Nevertheless, many kinematic dynamo simulations have been performed to optimize this flow,
using the synthetic MND flow \cite{MND06} or time-averaged flows measured in water-prototype experiments
\cite{MBD03,RCD05,SXG06}.

Different observations have been made that can help elucidate the exact mechanism 
by which the dynamo instability develops in the VKS experiment: 
i) in the exactly counter-rotating case, dynamo action is observed above a critical magnetic Reynolds 
number $R_m^c\approx 32$ and the dominant saturated mode is an axisymmetric  m=0 mode in cylindrical 
coordinates \cite{MBB07}; 
ii) so far, dynamo has only been observed with iron propellers \cite{P5}; 
iii) in a given experimental set up with iron disks, dynamo has been observed when counter-rotating 
disks in (+) 'non scooping' direction (with respect to the propellers blades), but no dynamo has 
been observed when counter-rotating the disks in the (-) 'scooping' direction \cite{P5}. 

Observation i) shows that the dynamo mechanism is not the simple  linear  response to the mean 
flow alone, since the latter is axisymmetric and produces a transverse $m=1$ mode ($m$ denotes the azimuthal wavenumber) 
above $R_m^c\simeq 43$ as indicated by kinematic dynamo simulations \cite{RCD05}. 
It has been suggested in \cite{GAFDFauve} that the m=0 
mode of the VKS experiment is in fact generated by non-axisymmetric fluctuations through an 
alpha-omega mechanism taking place within the blades of the propellers. This suggestion has 
been confirmed by recent numerical simulations \cite{LNRLGP08} showing that the addition of an 
(empirical and large) alpha term in the induction equation changes the dominant m=1 mode into a dominant m=0 
mode with a dynamo threshold close to the observed experimental critical value $R_m^c\approx 32$. Another 
interesting suggestion has been made by \cite{GDF08}, namely that the m=0 mode could 
arise from a secondary bifurcation of the m=1 mode, since the latter breaks the original axisymmetry
of the velocity field.

Observation ii) shows that magnetic boundary conditions have a large impact on the dynamo 
instability. High magnetic permeability conditions are known to decrease the dynamo threshold for the Roberts 
and Ponomarenko flows \cite{APG03} and more recently for von K\'arm\'an type flows \cite{GIFD08}. 
For iron disk, it is suggested that the dynamo threshold could be reduced by a 
factor as large as 2 with respect to stainless disks. Since it has not been yet possible to get beyond  $R_m=50$ in the VKS 
experiment, this could explain why dynamo has not been observed with stainless steel disks, while it is observed 
around $R_m=32$ with iron disks. Other kinematic simulations \cite{SXG06} have shown that the sodium flow behind the disks
can lead to an increase from $12\%$ to  $150\%$. Using iron disks can thus also screen magnetic effects of the flow behind 
the disks. PARLER des DISQUES CAMEMBERTS?

Observation iii) shows that the dynamo is not produced only by the rotation of magnetized iron disks in 
a conducting fluid but depends on the flow characteristics: it is a real fluid dynamo. 
However, not much has been done so far to try and exploit this observation. 
In particular, no kinematic dynamo simulations have been performed so far with experimental flows
including an annulus as in the VKS experiment or with propellers counter-rotating in the (-) 'scooping' direction.

In the present paper, we use Stereoscopic Particle Imaging Velocimetry (SPIV) measurements obtained in
von K\'arm\'an water (VKE) experiments modeling the VKS experiment to perform kinematic dynamo simulations. 
We investigate the role of the mean field structure on the dynamo threshold, 
with special attention to the rotation sense. Because of the axisymmetry, the exhibited kinematic mode is, 
as expected, different from the experimental saturated one. However, we show that the kinematic dynamo thresholds 
qualitatively vary like the observed dynamo thresholds and discuss how these results can be useful to understand 
the actual mechanisms at the origin of VKS dynamo.

Our paper is organized as follows: In Sec. II, after the problem formulation, we describe the experimental 
measurements and the numerical method. In Sec. III, we focus on the processing procedure of the measured 
velocities and perform a comparison with previous estimates based on Laser Doppler Velocimetry (LDV) measurements. 
Results for dynamo action are presented  in Sec. IV and compared  with the VKS experiments. 
Finally, we discuss the consequences of these results for the VKS dynamo.

\section{Methods}

\subsection{Problem Formulation}

In the present paper, we focus on the kinematic dynamo problem: we consider a given velocity field  
$\mathbf{u}$, e.g. as measured in a water experiment, and check its ability to sustain dynamo action 
by solving the magnetic induction equation
\begin{equation}
\frac{\partial \mathbf{b}}{\partial t}=\nabla
\times\left(\mathbf{u} \times
\mathbf{b}\right)+\frac{1}{\mu
\sigma}\triangle\mathbf{b},
\end{equation}
where $\mathbf{b}$ denotes the magnetic field with \mbox{$\nabla \cdot \mathbf{b}=0$}, $\mu$
and $\sigma$ are respectively the magnetic permeability and the electrical
conductivity of the considered medium, and $\Delta$ is the Laplacian operator. Since
$\mathbf{u}$ is given, the equation is linear in $\mathbf{b}$. There are two possible long-time
behaviour for the magnetic field: either it is decaying and there is no dynamo, or it is
increasing, and there is dynamo. Growth or decay is in general exponential (although there can be
transient algebraic growth, due to non-normality of the operator, see \cite{LJ06}). To quantify the
growth, we introduce the growth rate \mbox{$\sigma=\partial_t \ln\left<B^2\right>$} that is
based on the magnetic energy  \mbox{$\left<B^2\right>$}, where the
average is taken over the spatial domain.

Since the equation is linear, no saturation is observed in the kinematic approach. This
technique is therefore unable to detect subcritical or finite amplitude dynamos \cite{PLD07} 
or to study the saturation regime \cite{PMM05,LBL06,DBM07} as in numerical flows such as the 
Taylor-Green flow.CITER SIMULATIONS FOREST??.
However, this approach gives an estimation for the dynamo threshold as a function of the 
mean velocity field. 

\subsection{Experimental measurements}

\begin{figure}
\includegraphics[clip=true,width=0.99\columnwidth]{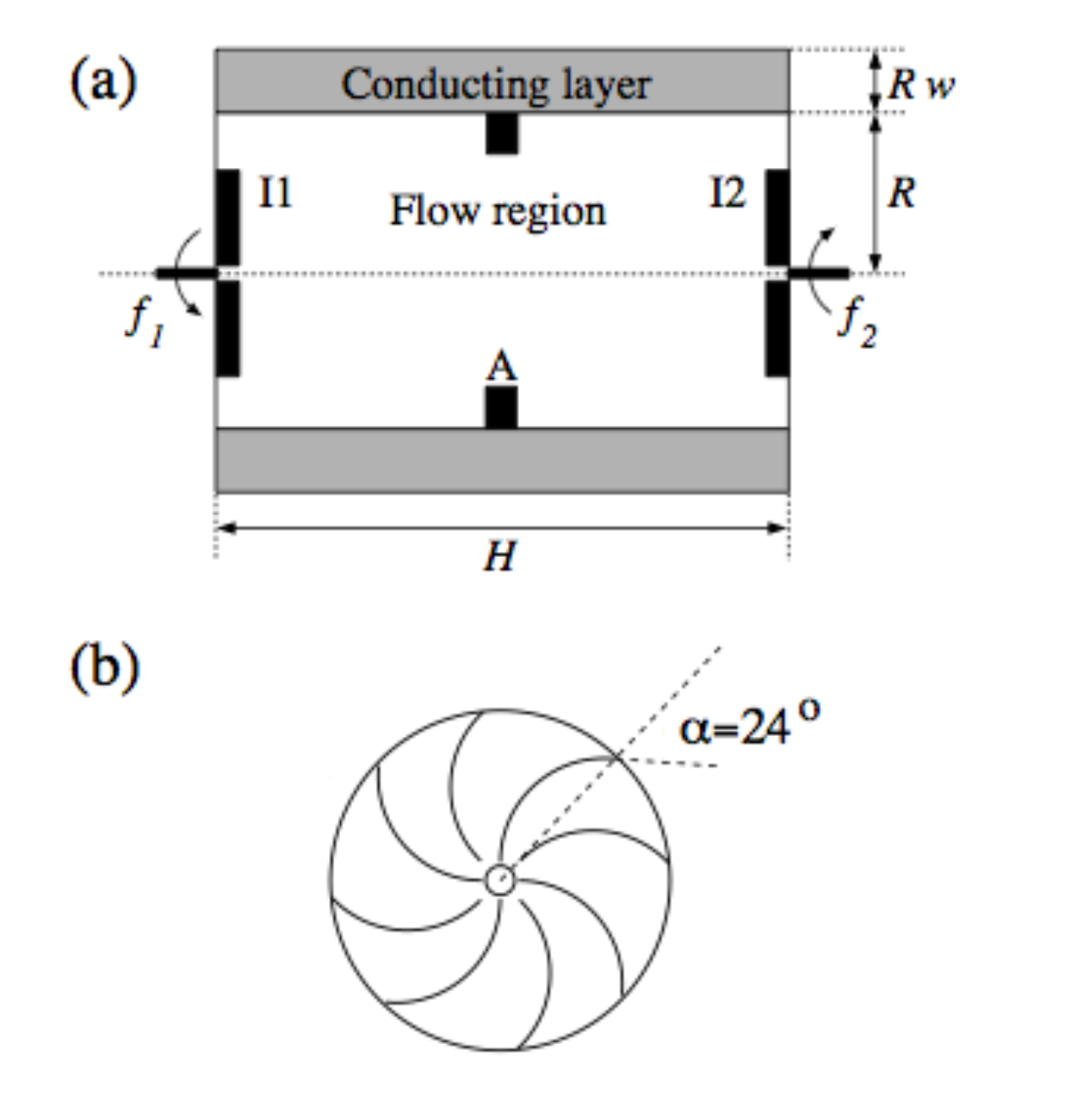}
\caption{(a) Sketch of the simulated part of the VKS experiment (cross
section) including both impellers (I1,I2) and an annulus (A) at mid height.
The horizontal axis corresponds to the $z$ axis. 
The container radius $R$ is taken as length scale. 
The factor $w$ denotes the thickness of the conducting layer (grey area) of sodium 
at rest in the VKS experiment. The layers of sodium between the impellers and the 
end walls of cylinder have not been taken into account in this work.
(b) Top view of the TM 73 impellers used in the VKS experiment \cite{RCD05}.
The (+) (resp. (-)) rotation sense corresponds to a counterclockwise (resp. clockwise)
rotating impeller.}
\label{FIG1}
\end{figure}

{\it Geometry and experimental setup.---}
The setup corresponding to the VKS experiment is described in details in \cite{MBB07} and
 sketched in Figure \ref{FIG1}(a). A cylindrical vessel with a fixed aspect ratio of $L=H/R=1.8$, where $R$
(resp. $H$) denotes the radius (reps. the length) of the vessel which is filled with liquid sodium.
We arrange the coordinate system such, that the axial extension of the cylinder is situated in the interval
 $z \in [-0.9,0.9]$. The fluid is stirred by two independent rotating so called TM73
impellors \cite{RCD05}, at frequencies $f_1$  and $f_2$. 
The impellers are fitted with curved blades (see Fig. \ref{FIG1} (b)). Therefore, the flow properties
depend on the sense of rotation, specifically on  whether  the blades are presenting their
concave ((-) rotation) or  convex ((+) rotation) side to the flow during impeller rotation. 
In the special case where  $f_1=f_2>0$ (resp. $f_1=f_2<0$) with the sign convention of Fig. \ref{FIG1}, the
impellers fullfill exact (+) (resp. (-)) counter-rotation. An optional concentric annulus can also be inserted 
at mid height and has been used in the first dynamo runs. This annulus reduces large scale  fluctuations 
\cite{CDM09} and can also have an impact on the mean flow, in particular on the toroidal to poloidal 
ratio $\Gamma$ (see below).  In the VKS experiment, an additional layer of resting sodium is present in 
the outer part of the experiment. 

The VKE experiment is the exact half-scale water model of the VKS experiment, except that the experiment is 
realized vertically (vertical z axis) whereas the VKS setup is horizontal. The cylinder radius and height of the VKE vessel 
are $R= 100$ and $240$ mm respectively. We have used TM73 impellers of diameter $150$ mm fitted with radial 
blades of height $20$ mm and curvature radius $92.5$ mm. The inner faces of the disks are $H=180$ mm apart. 
We have worked with different vessel geometries, allowing the insertion of an annulus - thickness $5$ mm, inner
diameter $170$ mm - in the equatorial plane. Note also that in the VKE experiment, the layer 
of resting fluid is absent. The velocity field is therefore only measured in the {\it flow region}, 
see Fig. \ref{FIG1} (a).\

\begin{figure}
\includegraphics[clip=true,width=0.99\columnwidth]{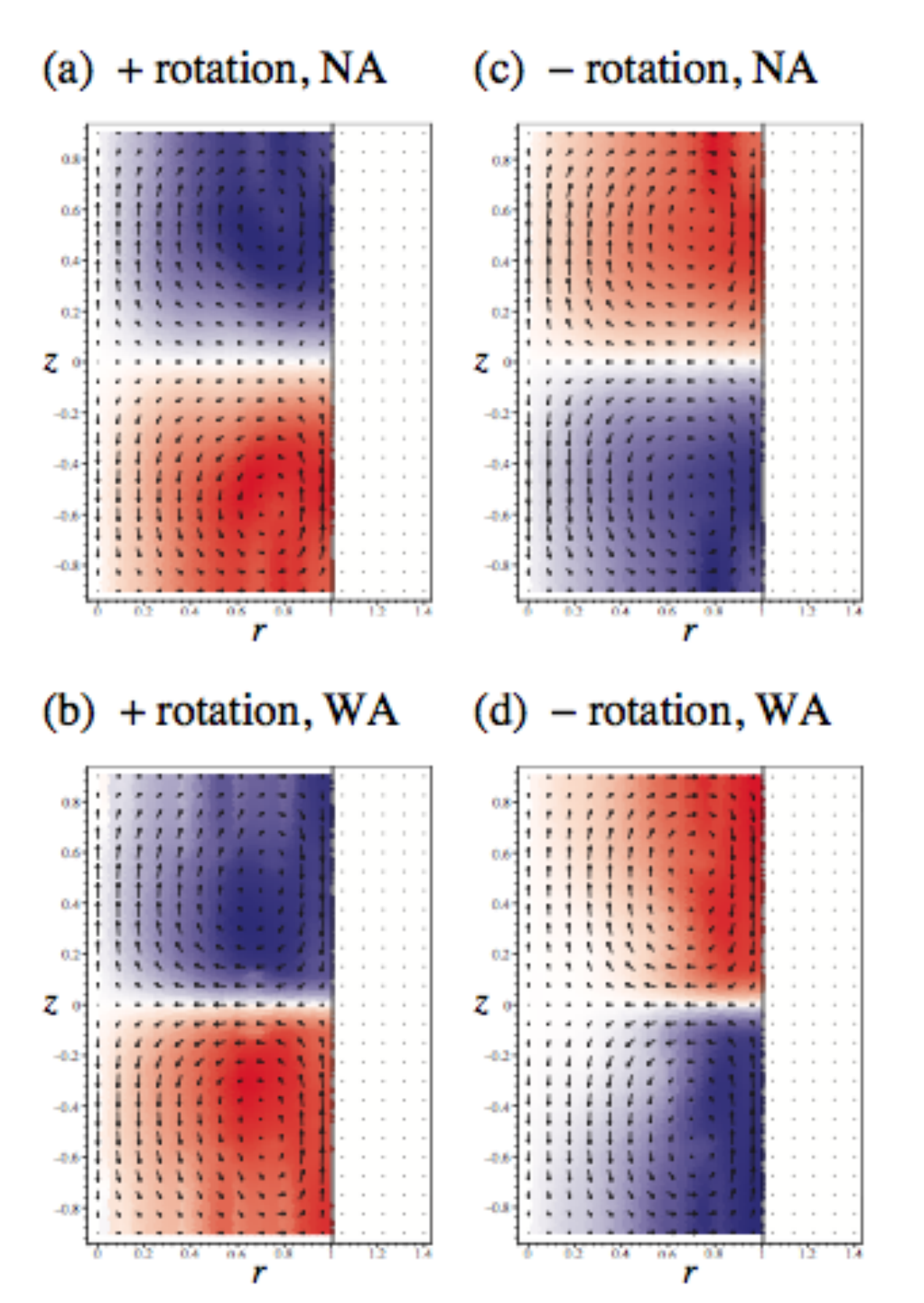}
\caption{Dimensionless velocity field measured by SPIV in VKE and symmetrized as well as 
spline extrapolated (nrp=4) for kinematic dynamo simulations (see text for details). 
(a) (+) rotation and no annulus (NA); 
(b) (+) rotation and presence of an annulus (WA); 
(c) (-) rotation and no annulus (NA); 
(d) (-) rotation and presence of an annulus (WA); 
The coloring displays the velocity field perpendicular to the plane, whereas red (blue)
denotes a positive (negative) sign in cylindrical coordinates $(r,\Theta,z)$.
The white area for radii $r>1$ shows the conducting layer at rest $(w=0.6)$
with $\mathbf{u}=0$.}
\label{FIG2}
\end{figure}

Velocity measurements are performed using a DANTEC SPIV system, providing the three
components of the velocity field on a plane through a $95 \times 66$ points grid \cite{M07}. 
In all that follows, the measurement plane is a meridian plane containing the rotation axis.
The velocity fields are averaged over time series of $5000$ regularly sampled values (sampling frequency 
between 1 and 15 Hz). In the exact counter-rotating regime, the system is axisymmetric and symmetric with 
respect to any rotation $R_\pi$ of $\pi$ around any radial axis in its equatorial plane. The mean flow 
is composed of two toroidal cells separated by an azimuthal shear layer and two poloidal recirculation cells.
Figure \ref{FIG2} shows the velocity fields obtained in the 4 studied configurations: (+) and (-) rotation sense 
in the presence or the absence of an annulus in the equatorial plane. Note that in the exact counter-rotating
regime, no turbulent bifurcation towards a one cell state is observed with the TM73 impellers contrary to what is
reported for impellers with higher curvarture blades in \cite{RMC04}. The bifurcation is observed in the (+) sense
for a rotation number $\theta= (f_1-f_2)/(f_1+f_2) = 0.1$ without annulus and $\theta = 0.17$ with annulus 
\cite{CDM09}.

{\it Control parameters.---}
Choosing the magnetic field diffusion time $\mu \sigma R^2$ as the time scale, and
the maximum absolute value of the flow velocity $|\mathbf{u}|_{\textrm{max}}$ as velocity unit,
the induction equation can be expressed in dimensionless form:
\begin{equation} \label{MHD}
\frac{\partial \mathbf{b}}{\partial t}=R_m \nabla
\times\left(\mathbf{u} \times
\mathbf{b}\right)+\triangle\mathbf{b},
\end{equation}
where \mbox{$R_m=\mu \sigma R |\mathbf{u}|_{\textrm{max}}$} is  the magnetic
Reynolds number. The critical magnetic Reynolds number $R_m^c$ for dynamo onset is then defined 
such that $\sigma(R_m^c)=0$. Another important parameter to characterize the
velocity field is the poloidal to toroidal ratio:
\begin{subequations} \label{Gamma}
\begin{eqnarray}
\Gamma&=&\frac{\left<P\right>}{\left<T\right>},\  \textrm{with}\\
\left<P\right>&=&\int_{-L/2}^{L/2} dz \int_0^1 
\sqrt{u_r(r,z)^2+u_z(r,z)^2} r dr,\\
\left<T\right>&=&\int_{-L/2}^{L/2} dz \int_0^1 \left|u_\Theta(r,z)\right| r dr,
\end{eqnarray}
\end{subequations}
which is known to have a great impact on the dynamo threshold \cite{MBD03,RCD05}. Here we denoted the components 
of the velocity field in cylindrical coordinates  as  $\mathbf{u}=(u_r,u_\Theta,u_z)$.

\subsection{Numerical code}
Because the system is axisymmetric, the time-averaged velocity field is supposed to be axisymmetric.
We integrate the induction equation using an axially periodic kinematic dynamo code.
We only provide here its brief description, details can be found in \cite{L94}. The code is
pseudospectral in the axial and azimuthal directions while the radial
dependence is treated by a high-order finite difference scheme. The numerical
resolution corresponds to a grid of $81$ points for one wavelength in axial direction, $4$ points
in azimuthal direction (corresponding to azimuthal wave numbers $m=0,\pm 1$) and $31$
points in radial direction for the flow domain. The time scheme is second-order
Adams-Bashforth with a typical time step of $10^{-5}$.

Electrical conductivity and magnetic permeability are homogeneous and the external
medium is insulating. Implementation of the magnetic boundary conditions for a finite
cylinder is difficult, due to the nonlocal character of the continuity conditions at
the boundary of the conducting fluid \cite{GIFD08}. In contrast, axially periodic boundary conditions
are easily formulated, since the harmonic external field then has an analytical expression.
Indeed, the numerical elementary box consists of two mirror-symmetric experimental
velocity fields in order to avoid strong velocity discontinuities along the $z$ axis
\cite{RCD05}. Note that we do not take into account the effet of the flow behind the propellers. 
This effect has been studied in details in other kinematic simulations \cite{SXG06}.

Finally, we can act on the electromagnetic boundary conditions by adding a layer of
stationary conductor of dimensionless thickness $w$, corresponding to sodium at rest
surrounding the flow exactly as in the VKS experiment in Cadarache \cite{MBB07}. 
This extension is made keeping the radial resolution of the experimental
measured velocity field constant ($31$ points in the flow region). The velocity field we
use as input for the numerical simulations is thus simply fit in an homogeneous
conducting cylinder of radius $1+w$:
\begin{subequations} \label{conductlayer}
\begin{eqnarray}
\mathbf{u}&=&\mathbf{u}_{\textrm{flow}} \quad
\textrm{for} \quad 0\leq r \leq 1,\\
\mathbf{u}&=&0 \quad \textrm{for} \quad 1 < r \leq 1+w.
\end{eqnarray}
\end{subequations}
Former investigations \cite{RCD05} showed a decrease of the critical magnetic Reynolds number
for increasing conducting layer size,  with a saturation above \mbox{$w=0.4$}. Therefore, we
fixed \mbox{$w=0.4$} (meaning altogehter $43$ points in radial direction) in order to save computing time.
Some calculations with \mbox{$w=0.6$} ($49$ points) have also been done, when comparing with
older computations based on LDV measurements.

\begin{figure}
\includegraphics[clip=true,width=0.99\columnwidth]{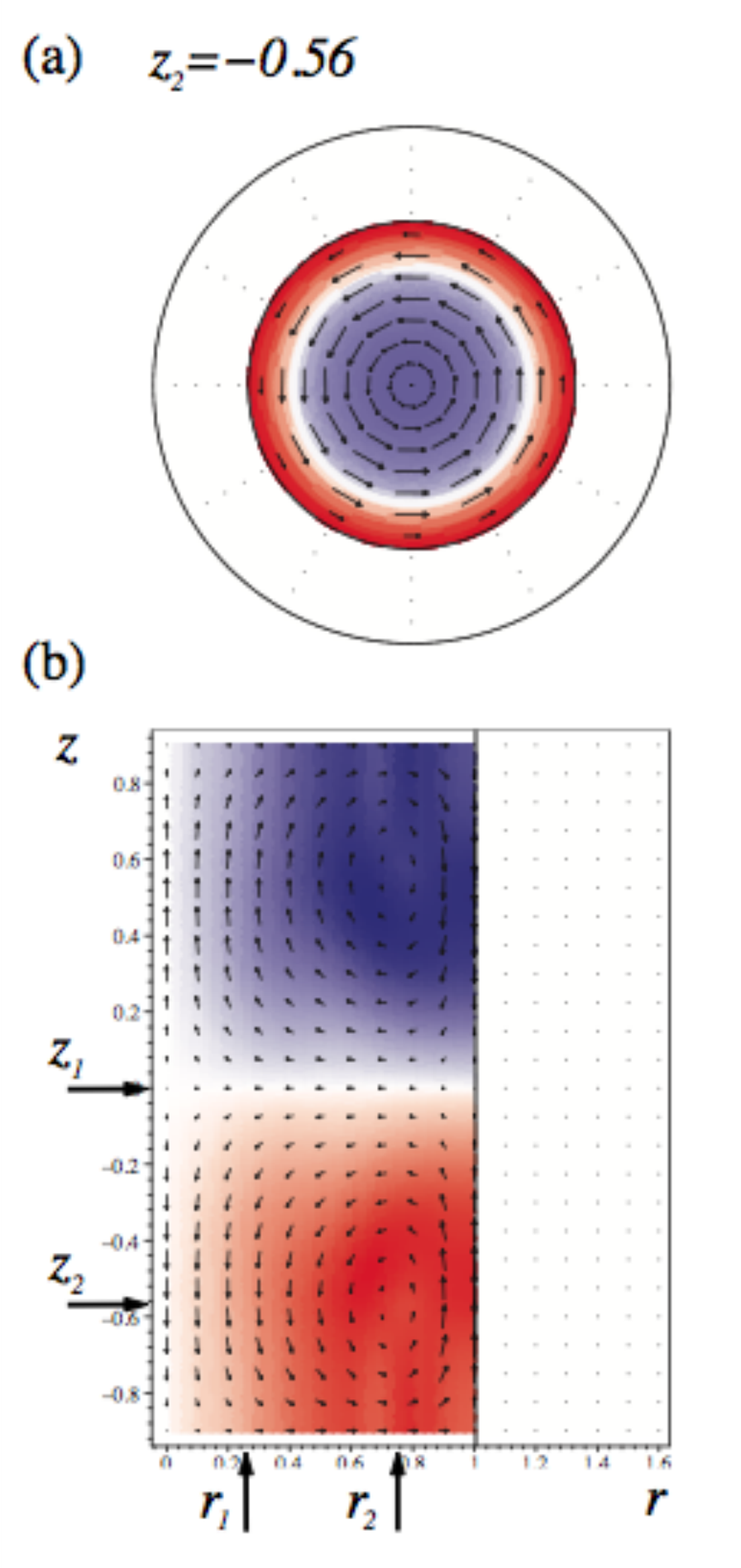}
\caption{Same as in Figure \ref{FIG2}: (+) rotation has been used and there is no annulus.
Part (a) indicates a section for fixed $z=z_2=-0.56$, whereas in (b) an arbitrary cut of
the rotational invariant field for fixed $\Theta$ is displayed. The arrows in (b)
illustrate the section planes for the profiles shown in figures \ref{FIG4}
and \ref{FIG5}.}
\label{FIG3}
\end{figure}

Figure \ref{FIG3} gives an example of the total velocity field used in kinematic simulations,
consisting of an experimentally measured velocity field  $\mathbf{u}_{\textrm{flow}}$
and a conducting layer with \mbox{$w=0.6$}.

\section{Data preparation and tests}

\subsection{Preparation of measured velocities\label{Preparation-method}}

Due to experimental measurement constraints, it is impossible to determine the
field in the whole cylinder cut \mbox{$(r,z)=\left([0,1],[-0.9,0.9]\right)$}:
i) in the propeller regions, the presence of blades prevents the SPIV measurements; 
ii) close to the cylinder, reflection effects spoil any velocity measurements; 
iii) when an annulus is inserted at mid-height ($z=0$), measurements cannot be 
performed in the thin area behind and around the annulus.
Therefore a typical data set covers only the area
\mbox{$(r,z) \approx \left([0,0.9],[-0.6,0.6]\right)$}. This means,
that approximately more than a third of the velocity field is unknown and has to be
determined by boundary conditions and spline extrapolation. The radial adaptation has
always been performed before the axial one. The preparation methods of the used SPIV
measurements are explained in detail in the following.

{\it Axial Symmetrization.---} For the perfect counter-rotating setup an axial
symmetrization has been performed with height mid as reflection plane.
After this procedure $u_r$ is strictly symmetric, whereas $u_\Theta$ and $u_z$
are antisymmetric.

{\it Axial boundary conditions.---} The axial velocity field should vanish at the blade border,
therefore we imposed the condition $u_z(z=\pm 0.9)=0$. The two other
components fullfill a vanishing derivative in axial direction at the blades, meaning 
\mbox{$\partial u_\Theta/\partial z(z=\pm
0.9)=0$} and \mbox{$\partial u_r/\partial z(z=\pm 0.9)=0$}.

{\it Extrapolation in axial direction.---} Close to the blades the fields are unknown and even not
all the data which are measured are suitable, meaning that inappropriate
values have to be removed. This lack is repaired by a spline extrapolation.
The aforementioned boundary conditions help us to implement a good fit. The procedure at the top
and the bottom blade is the same. The number of points which have to be
removed at a blade are denoted by nrp (=``number of removed points'') in the following
text. We compared results for different values of nrp in order to find a
trustable result with a respective error estimate. Altogether between 11 and 15 points
have to be extrapolated at a blade corresponding to a nrp between $2$ and $6$,
meaning that for the first $9$ points no measured values are available. The
number of points used to determine the spline function has no significant influence on the
result, as far as more than $3$ are taken into account. For security reasons we always took $6$ points.

{\it Interpolation in the annulus region.---}
When the annulus is present, there are typically $8$ (10~\% of the data set) measured points
around the mid-height that are absent or spoiled. They have been removed and replaced by spline
interpolated values, which is a good natured operation.

{\it Radial boundary conditions.---} The radial velocity field vanishes at the cylinder border,
therefore we imposed the condition $u_r(r=1)=0$. The two other components
satisfy \mbox{$\partial u_\Theta/\partial r(r=1)=0$} and \mbox{$\partial
u_z/\partial r(r=1)=0$}. In the cylinder center the radial and azimuthal components are zero,
meaning $u_r(r=0)=u_\Theta(r=0)=0$, whereas the axial field has a disappearing derivative
\mbox{$\partial u_z/\partial r(r=0)=0$}.

{\it Extrapolation in radial direction.---} Close to the cylinder the last two points have been
replaced and an additional point has been added to reach the cylinder
border. They are the result of a spline extrapolation fullfilling the
aforementioned boundary conditions using two points for the function construction.

{\it Check with LDV measurements.---}
LDV measurements are free of certain deficiency of the SPIV measurements, especially near the cylinder
boundary, where reliable velocity fields can be obtained. It is therefore interesting to test the
validity of our SPIV field preparation by comparison with LDV fields obtained previously by
Ravelet et al. \cite{RCD05}. Figure \ref{FIG4} compares the velocity fields of
LDV with the new SPIV measurements for fixed radial coordinate cuts 
explained in Fig. \ref{FIG3}. One cut is close to the vortex
center, the other one far away. A setup with (+) counter rotation has been used
without annulus. In addition, the processed data, used in the code, with the
aid of symmetrization and spline extrapolation for different nrp are shown. The difference
between the two curves is mainly in the azimuthal field $u_\Theta$ nearby
the blades. Note, that the large deviations of the two processed curves for the $u_z$ field in
the regime $r \approx [-0.6,0.6]$ is only due to the performed scaling described in the caption.
The SPIV dataset is finer and overall the comparison is quite satisfactory.

Corresponding cuts for fixed axial coordinate are shown in Fig. \ref{FIG5} for planes in the vortex
region and in the pure inflow layer separating the vortex pair. The agreement between LDV and SPIV 
measurements is good even if a small deviation can be observed on $u_r$.

AJOUTER PHRASE CONCLUSION: AVANTAGE SPIV/LDV

\subsection{Influence of the annulus}

In order to study the influence of the annulus, we have measured the corresponding time-averaged velocity 
field for different nrp (cf. Fig. \ref{FIG6}). The velocity fields corresponding to the 4 configurations 
that are used in the kinematic dynamo simulations are then shown in Figures \ref{FIG7} and \ref{FIG8}.
 
In the (+) rotation case, the axial gradient of $u_r$, $u_\theta$ and $u_z$ are larger in the region of the shear layer.
The variation with $r$ of the velocity fields looks very similar except for $u_r$ at mid-radius. 
However, the poloidal to toroidal ratio $\Gamma$, does not reflect this difference: $\Gamma = 0.8$ in both cases.\

In the (-) rotation case, we were not able to perform such a detailed comparison, due to an
unexpected experimental problem: the locking of large scale vortices of the middle shear layer at
the position of the small cut performed through the annulus to enable SPIV
measurements (cf. \cite{CDM09} for details). This locking does not occur 
in the (+) case, resulting in an axisymmetric time-averaged velocity field. 
In the (-) rotating case, however, the locking artificially breaks the axisymmetry 
of the time-averaged field, preventing a priori the comparison with the case without 
annulus. Computation of  $\Gamma$  reveals a difference between the left ($\Gamma= 0.89$) 
part of the SPIV plane where the vortices are trapped and the right ($\Gamma= 0.62$)
part where there is no trapping. Other measurements show that only this right part is relevant 
for the VKS experiment in which the annulus is cut-free.  In the following, we use this right part 
and symmetrize the velocity field accordingly. The corresponding $\Gamma$ in (-) rotation sense 
are thus $0.62$ with annulus, and $0.49$ without annulus.\

To see the influence of $\Gamma$ on dynamo action for these velocity fields, we have also built a 
synthetic velocity field by taking the velocity field measured in the (-) rotation case without annulus, 
and rescaling its poloidal components to reach $\Gamma = 0.8$. This synthetic field is represented in 
Fig. \ref{FIG7} and \ref{FIG8}. The variations of its velocity components with $z$ and $r$
are very similar to those obtained for the (+), contary to the variation of the (-) case with annulus.

\subsection{Test of the code and the preparation method}

Several kinds of tests have been performed in order to check the numerical code as well as the
preparation methods of the SPIV data described in section \ref{Preparation-method}.

{\it Ohmic diffusion.---} 
The magnetohydrodynamic equations (\ref{MHD}) reduce for vanishing
velocity field $(\mathbf{u}=0)$ to the ohmic diffusion equation in cylindrical geometry
\begin{equation} \label{OHMIC} \frac{\partial \mathbf{b}}{\partial t}=\triangle\mathbf{b},
\end{equation}
which has analytical eigenfunctions consisting of Bessel functions \cite{MND06}. We imposed these
eigenfunctions in the code with a subsequent monitoring of the exponential
decay of the amplitude. In all the investigated cases the theoretical eigenvalue
has also been observed in the numerical run with high accuracy.

{\it Comparison with analytical flow results.---}
We used also the MND flow, an analytical velocity field studied in \cite{MND06} in order to ckeck
our code. For the field given by:
\begin{subequations} \label{analytfield}
\begin{eqnarray}
u_r&=&-\frac{\pi}{2}r(1-r)^2(1+2r)\cos(\pi z),\\
u_\Theta&=&(1-r)(1+r-5r^2)\sin(\pi z/2),\\
u_z&=&4\epsilon r(1-r)\sin(\pi z),
\end{eqnarray}
\end{subequations}
with aspect ratio $L=2$, for \mbox{$\Gamma=0.776$} and \mbox{$\epsilon=0.743$}, the expected critical
magnetic Reynolds number is $58$, which we could confirm with our code.

\begin{figure}
\includegraphics[clip=true,width=0.99\columnwidth]{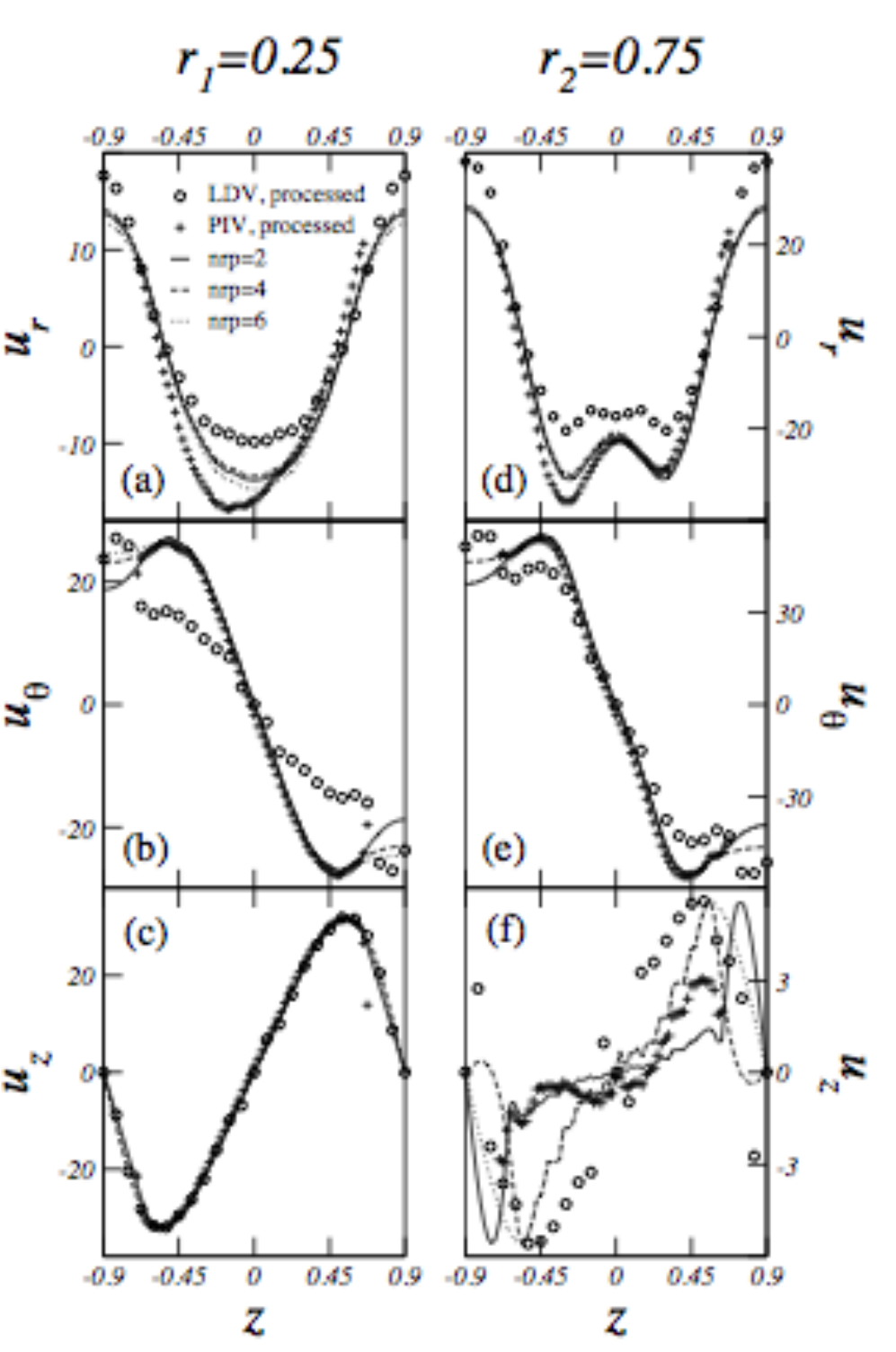}
\caption{Profiles of the velocity fields in cylindrical coordinates
$\mathbf{u}=(u_r,u_\Theta,u_z)$ for fixed radii as explained in Fig.
\ref{FIG3}. A setup with (+) rotation has been used and  there is no annulus. 
The left (right) column shows a section far away (nearby) the vortex
center. Data obtained by processed LDV \cite{RCD05} (circles) are compared with SPIV
measurements (crosses). The lines show processed SPIV data used in the kinematic
dynamo code for different spline extrapolation procedures, see text for details. 
All fields are in arbitrary units and scaled to the maximal amplitude of the LDV 
data for comparison reasons. Note that in part (f) the the scaling is expanded and 
the SPIV data values are very small.}
\label{FIG4}
\end{figure}

\begin{figure}
\includegraphics[clip=true,width=0.99\columnwidth]{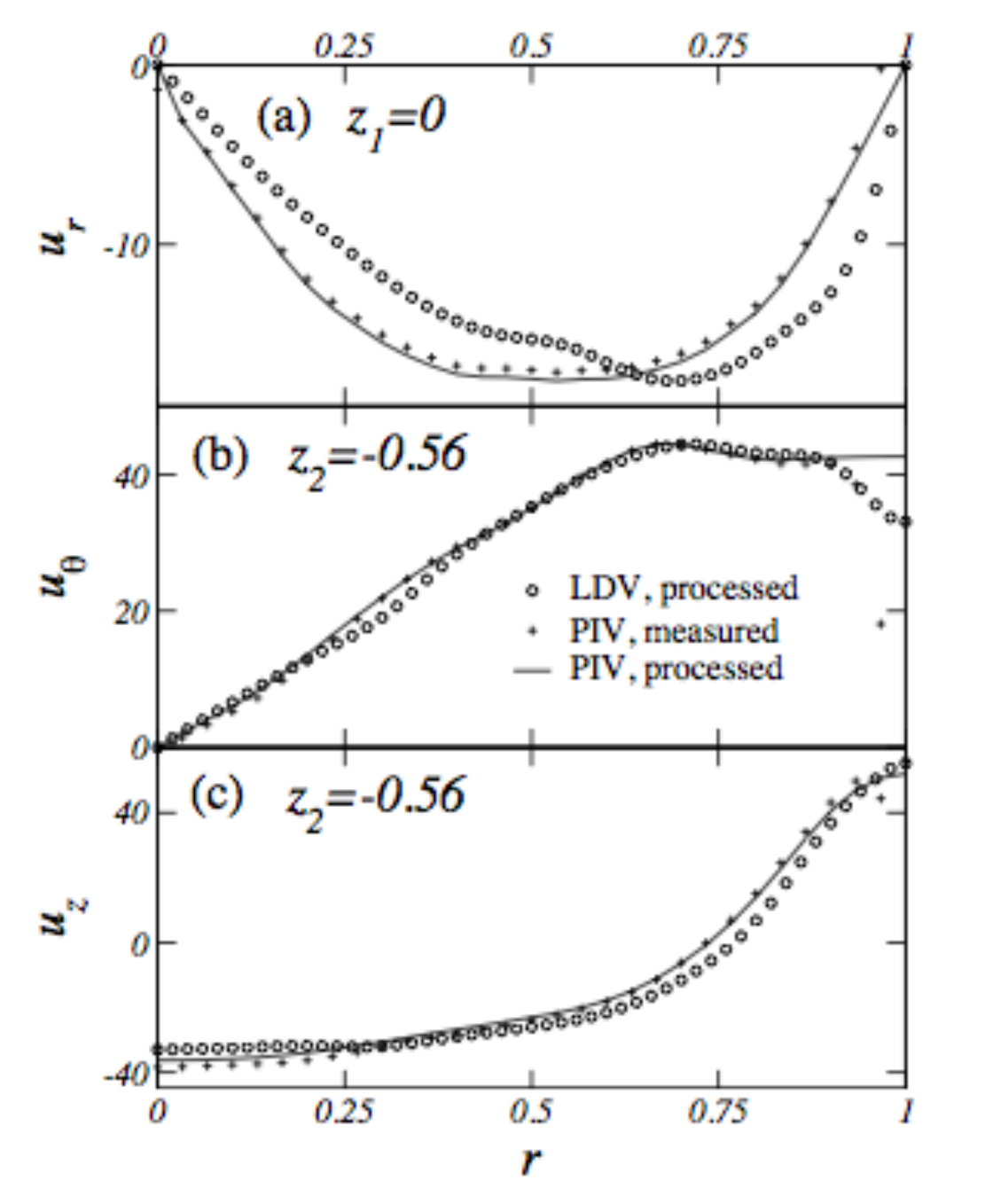}
\caption{Profiles of the velocity fields in cylindrical coordinates
$\mathbf{u}=(u_r,u_\Theta,u_z)$ for fixed axial heights as explained in Fig.
\ref{FIG3} and in caption of Fig. \ref{FIG4}. (+) rotation has been used 
and there is no annulus. For these cuts (far away from the blades)  
there is no difference in the different spline extrapolations. Fields are shown
in planes capturing their maximal value: (a) $u_r$ in the inflow area
separating the vortex pair at $z=z_1=0$; (b) and (c) $u_{\Theta}$ and $u_z$ in 
the vortex center at $z=z_2=-0.56$. Data obtained by processed LDV \cite{RCD05} 
(circles) are compared with SPIV measurements (crosses) and processed data (line).}
\label{FIG5}
\end{figure}

\begin{figure}
\includegraphics[clip=true,width=0.99\columnwidth]{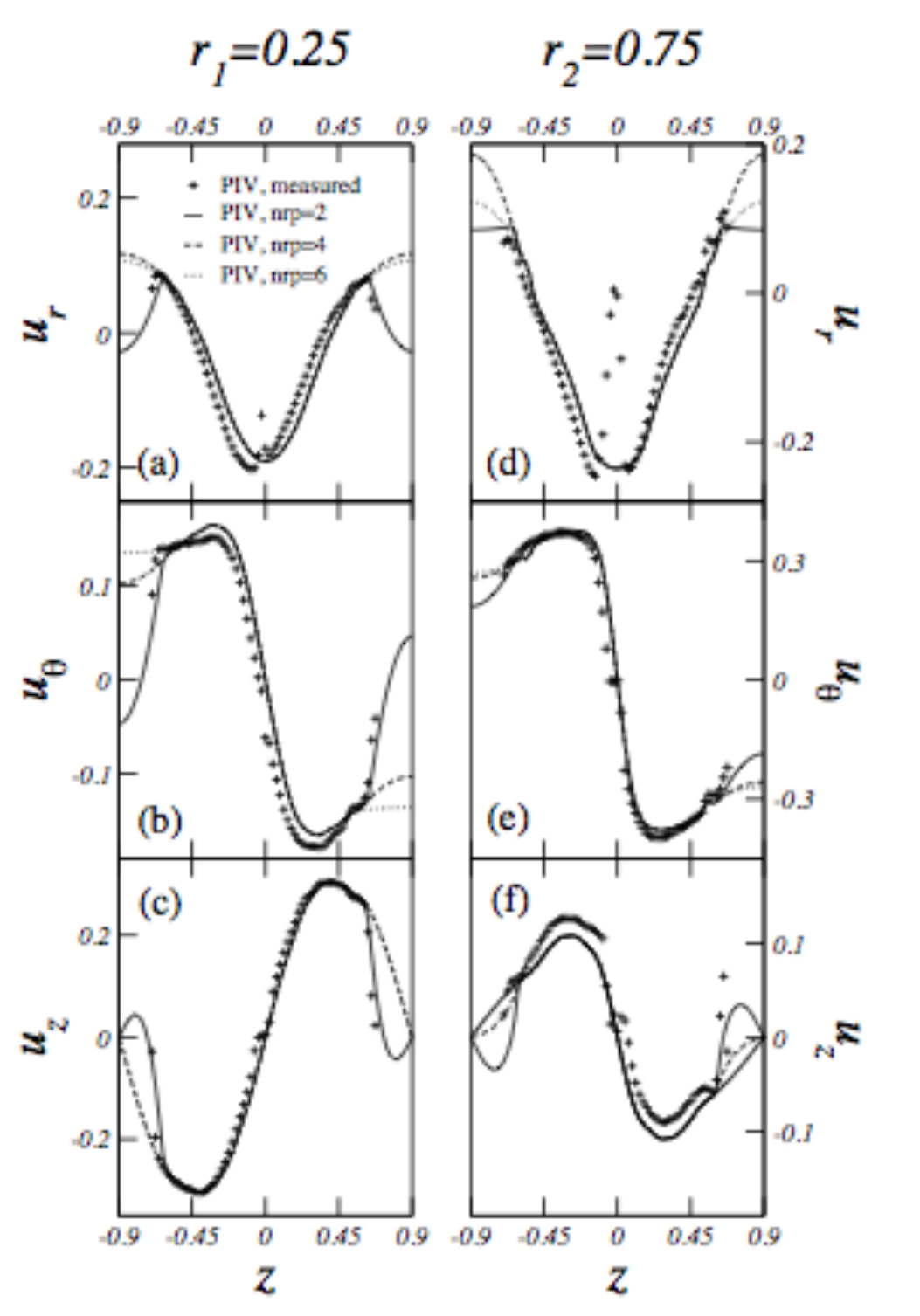}
\caption{Same as in Fig. \ref{FIG4} for a setup with (+) rotation and the presence 
of an annulus at mid height.}
\label{FIG6}
\end{figure}

\begin{figure}
\includegraphics[clip=true,width=0.99\columnwidth]{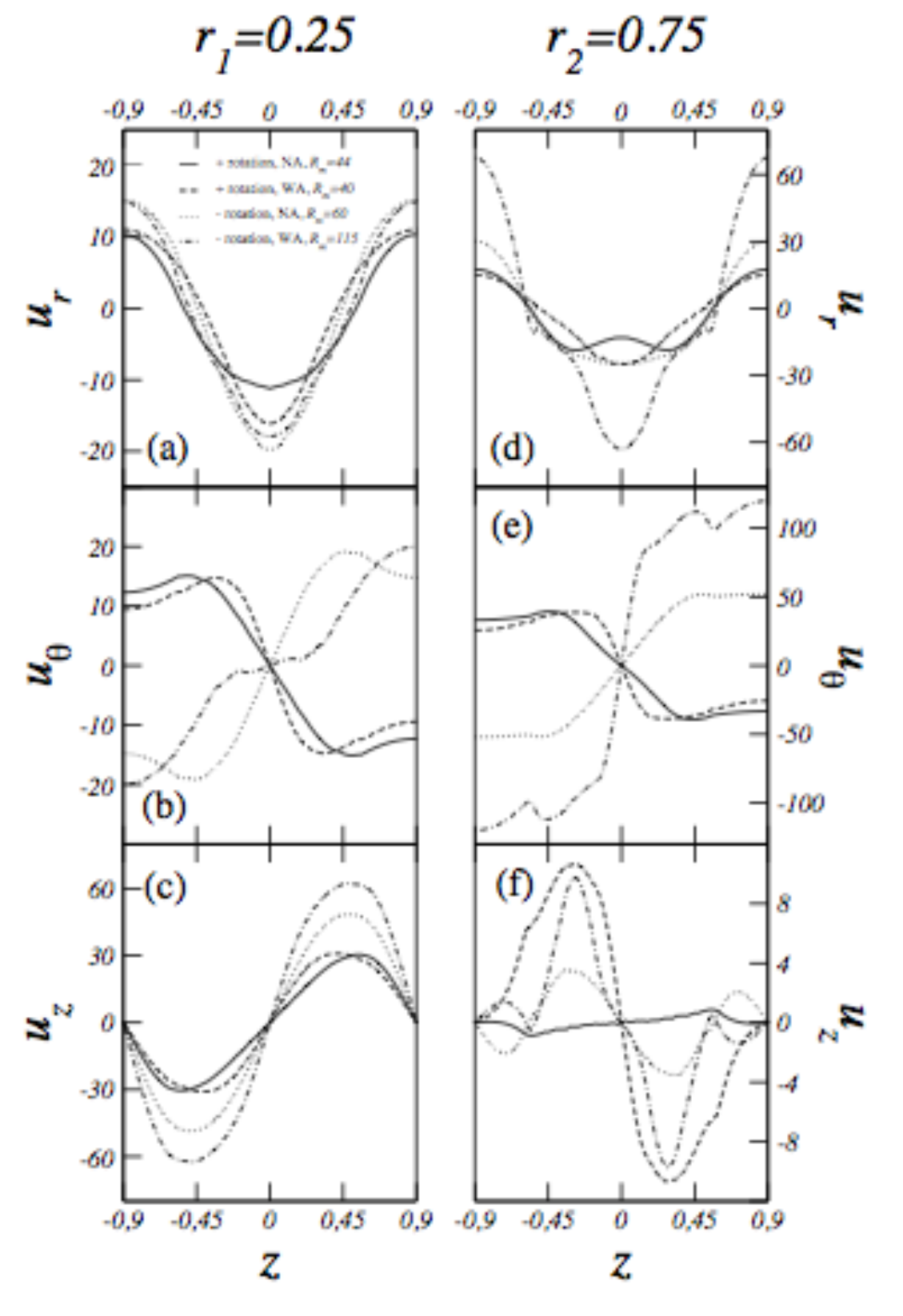}
\caption{Comparison of the profiles of the velocity fields corresponding to 
the 4 studied configurations for fixed radii as explained in Fig. \ref{FIG3}: 
(+) rotation without annulus (continuous line); (+) rotation with annulus at mid height (dashed line); 
(-) rotation without annulus: artificial numerical field with $\Gamma=0.8$ (dotted line); 
(-) rotation with annulus at mid height (dashed-dotted line)}
\label{FIG7}
\end{figure}

\begin{figure}
\includegraphics[clip=true,width=0.99\columnwidth]{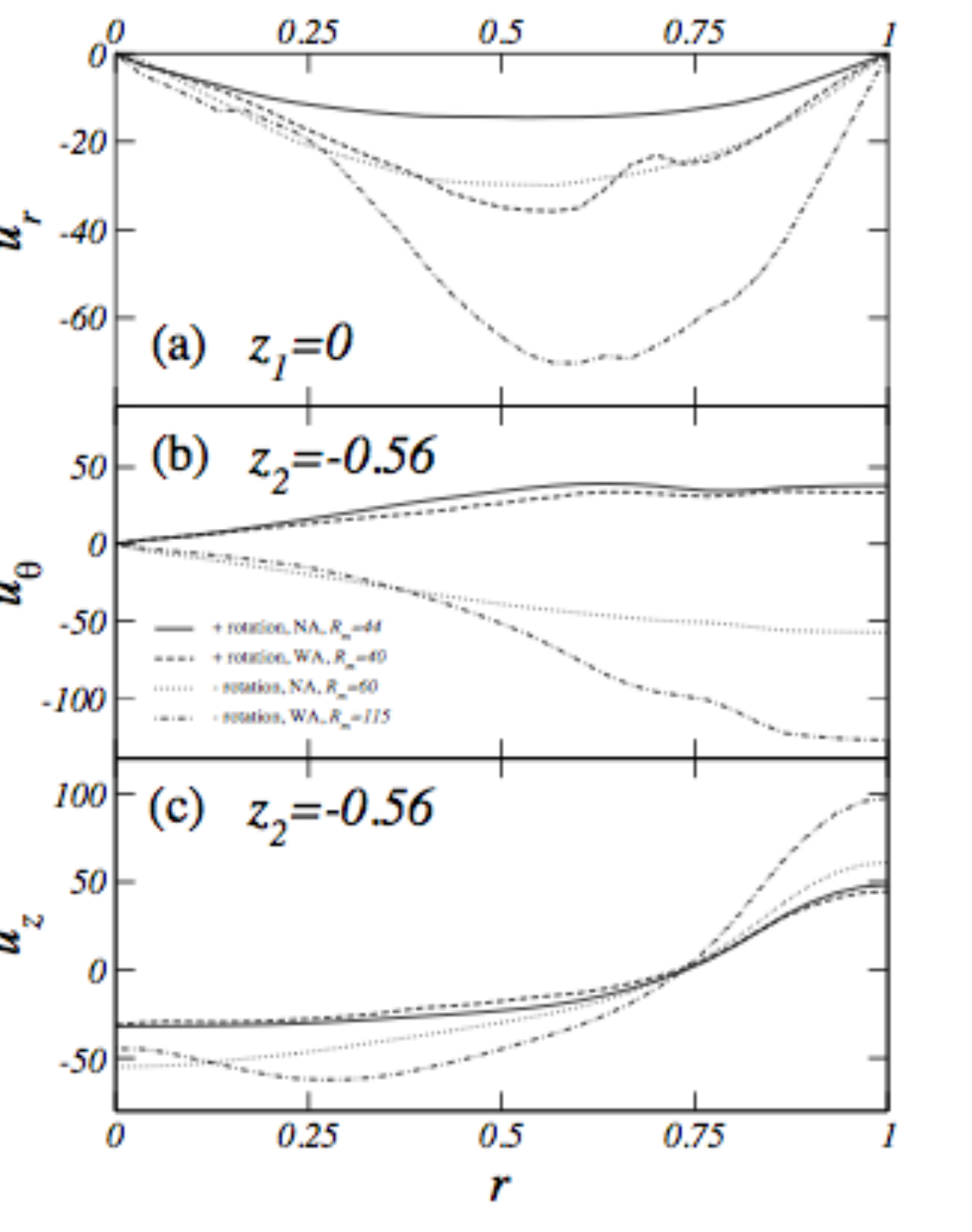}
\caption{Comparison of the profiles of the velocity fields corresponding to 
the 4 studied configurations for fixed axial height as explained in Fig. \ref{FIG3}: 
(+) rotation without annulus (continuous line); (+) rotation with annulus at mid height (dashed line); 
(-) rotation without annulus:: artificial numerical field with $\Gamma=0.8$ (dotted line); (-) rotation with annulus at mid height (dashed-dotted line)}
\label{FIG8}
\end{figure}

\begin{figure}
\includegraphics[clip=true,width=0.99\columnwidth]{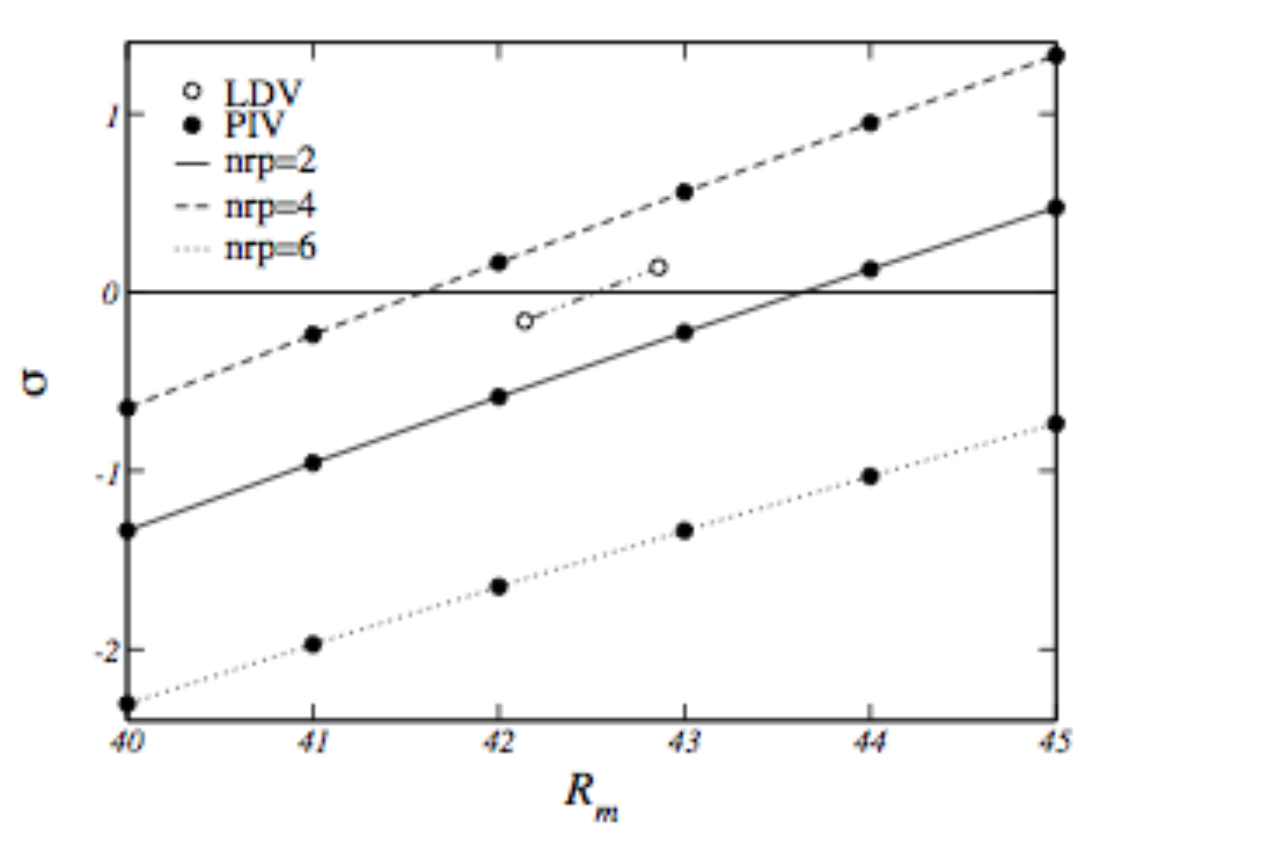}
\caption{Magnetic energy growth rate $\sigma$ vs. $R_m$ for a setup with a conducting layer
$(w=0.4)$ and a flow obtained with (+) rotation and without annulus, $\Gamma=0.8$. Full (open)
symbols refer to SPIV (LDV) data. For SPIV the results for different spline fits are compared,
cf. figures \ref{FIG4} and \ref{FIG5}, respectively.}
\label{FIG9}
\end{figure}

{\it Comparison with LDV data.---}
In order to check our preparation procedure of the SPIV measurements we compared also our
computations with former LDV results presented in \cite{RCD05}. The used fields have already been
presented in figures \ref{FIG4} and \ref{FIG5} for a (+) rotating setup
without annulus. Therefore we computed for different nrp values, namely $2,4$ and $6$, 
the magnetic energy growth rate as a function of $R_m$. The results are presented in 
Fig. \ref{FIG9} and compared with the corresponding LVD curve. The computed critical magnetic
Reynolds number of SPIV varies in a range $R_m^c = 41.6-47.6$, the highest value
corresponding to nrp=6. The comparison of velocity profiles in Fig. \ref{FIG4}, shows that, 
for nrp=6, too many points have been cut. The respective profiles are not very different and a 
significant difference is only observed in the $u_\Theta$ field, close to the impellor. 
This small difference is thus able to change the onset about 15\%. All these values are close to 
the LDV onset of $R_m^c = 42.5$. 

\begin{figure}
\includegraphics[clip=true,width=0.99\columnwidth]{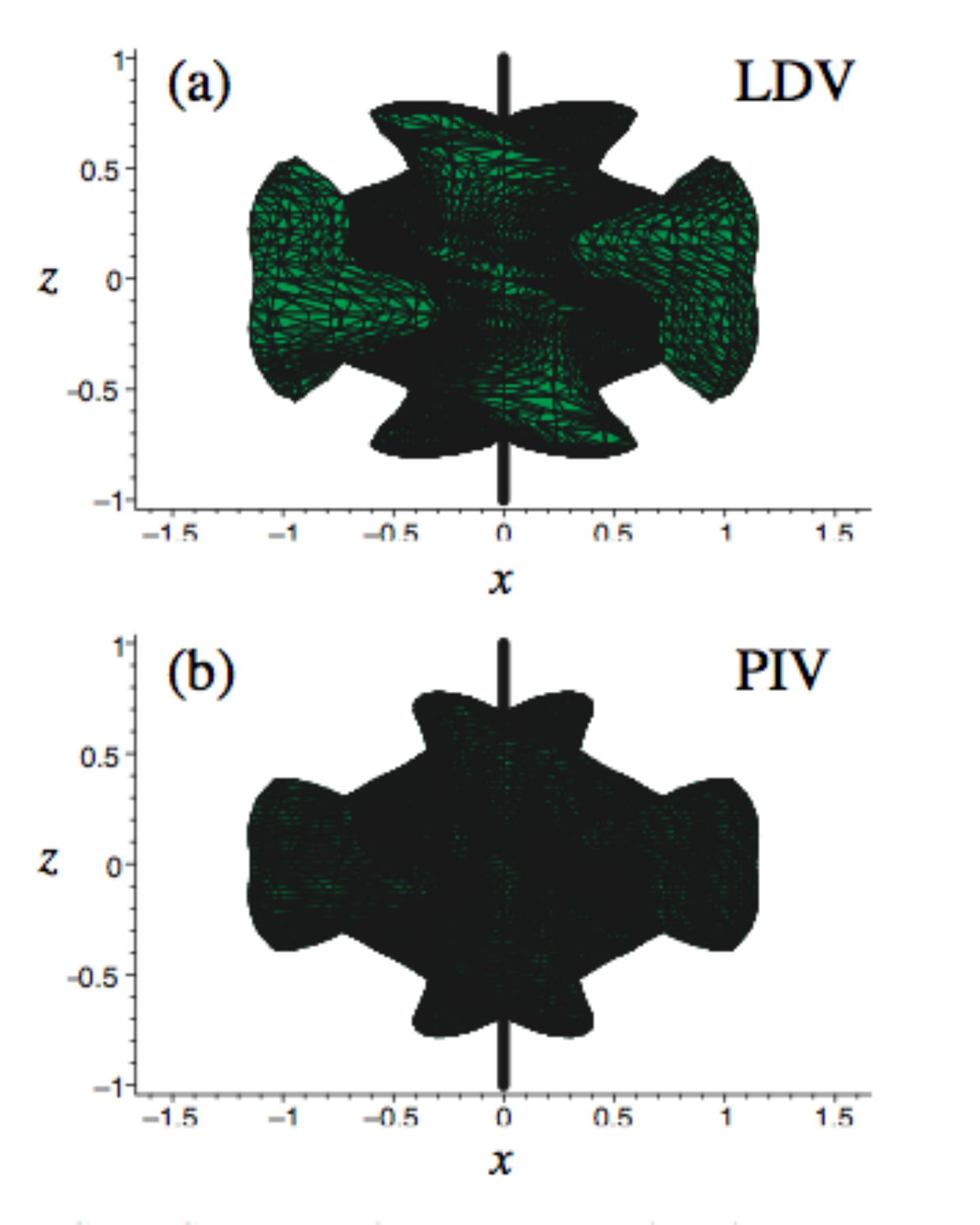}
\caption{Comparison of the isodensity surface of magnetic energy ($50 \%$ of the maximum) for LDV
(a) and SPIV (b). The figures show the neutral mode close to onset
$(R_m=44)$ with conducting layer $(w=0.6)$. A setup with (+) rotation
and without annulus has been used. For SPIV the spline extrapolation with nrp=4 was used.
The full line indicates the cylinder axis. Note, for $x>0$ (resp. $x<0$) the angle $\Theta$ is $0$ (resp. $\pi$).}
\label{FIG10}
\end{figure}

In Fig. \ref{FIG10} we furthermore compare the isodensity surfaces of the magnetic energy of two
runs for LDV and SPIV data with identical parameters. SPIV calculations have been done on a much finer grid in
axial direction. Both facings look very similar, which is confirmed by quantitative measurements.

\section{Kinematic dynamo simulations}

\subsection{Results}
\begin{figure}
\includegraphics[clip=true,width=0.99\columnwidth]{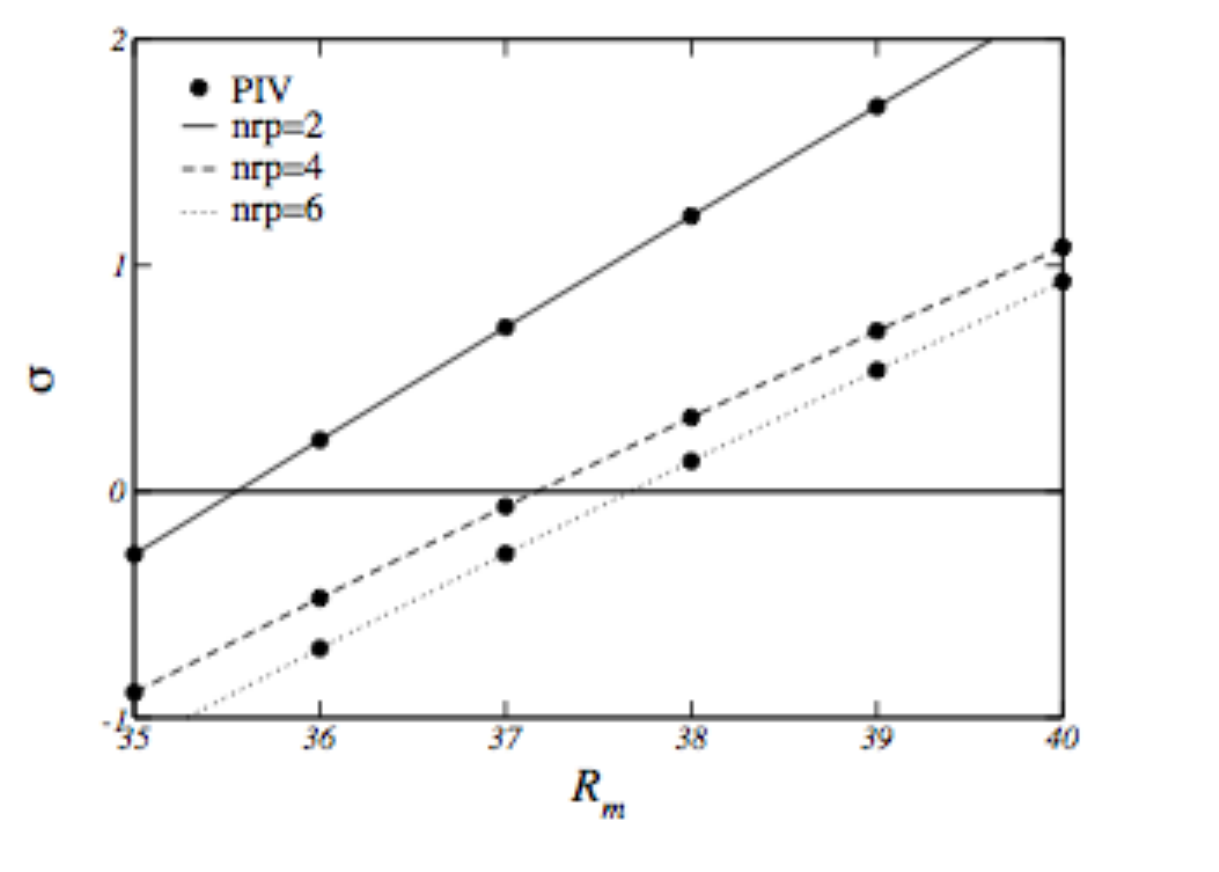}
\caption{Same as in Fig. \ref{FIG9} for a setup with (+) rotation and the presence 
of an annulus at mid height, $\Gamma=0.8$.}
\label{FIG11}
\end{figure}

\begin{figure}
\includegraphics[clip=true,width=0.99\columnwidth]{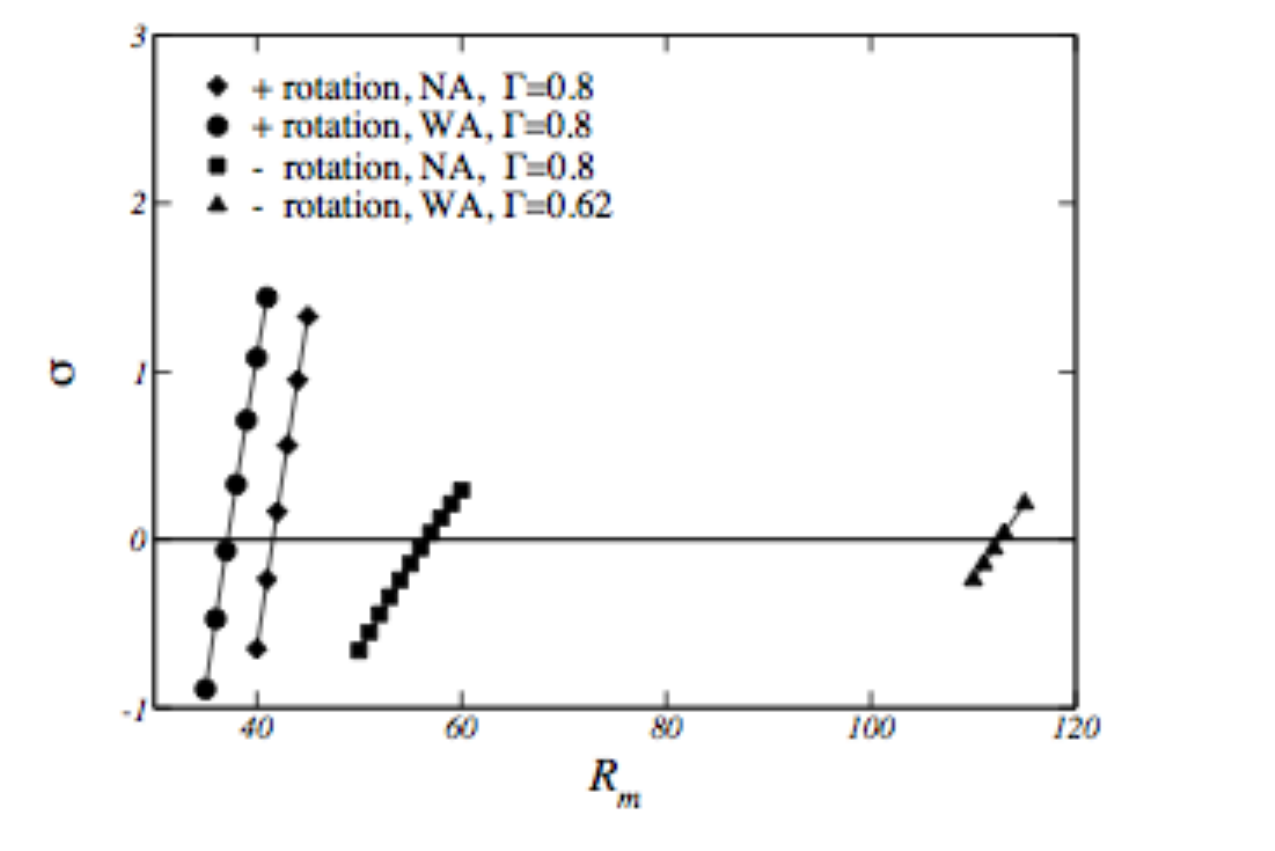}
\caption{Magnetic energy growth rate $\sigma$ vs. $R_m$ for different setups with
conducting layer $(w=0.4)$. WA (resp. NA) denotes configurations with (resp. without) annulus.
For all cases nrp=4 has been used. Note that the case (-) rotation without annulus corresponds to an artificial 
numerical field with $\Gamma=0.8$ not available in experiments.}
\label{FIG12}
\end{figure}

{\it Onset of dynamo action.---}
We have investigated the configurations in which the two propellers are exactly counter-rotating, in
(+) and in  (-) direction in setups with and without annulus. The computations have been made with nrp=2 to 6, and the
critical Reynolds number is found by computing the growth rate for different magnetic reynolds
numbers and the condition $\sigma(R_m^c)=0$. The results are presented in Fig. \ref{FIG11} and \ref{FIG12} and
summerized in  Table \ref{TAB1}.
 
In the presence of an annulus in the shear layer, the results show that the direction of rotation changes
the critical magnetic Reynolds number by a factor 3 :  $R_m^c \simeq 36$ for the (+) case and $R_m^c \simeq 113$ 
for the (-) case. There is an ever larger difference between the (+) and (-) rotation case when the annulus is absent:
the dynamo threshold is $R_m^c \simeq 45$ for the (+) case, and  $R_m^c > 500$ in the (-) case. 

\begin{figure}
\includegraphics[clip=true,width=0.99\columnwidth]{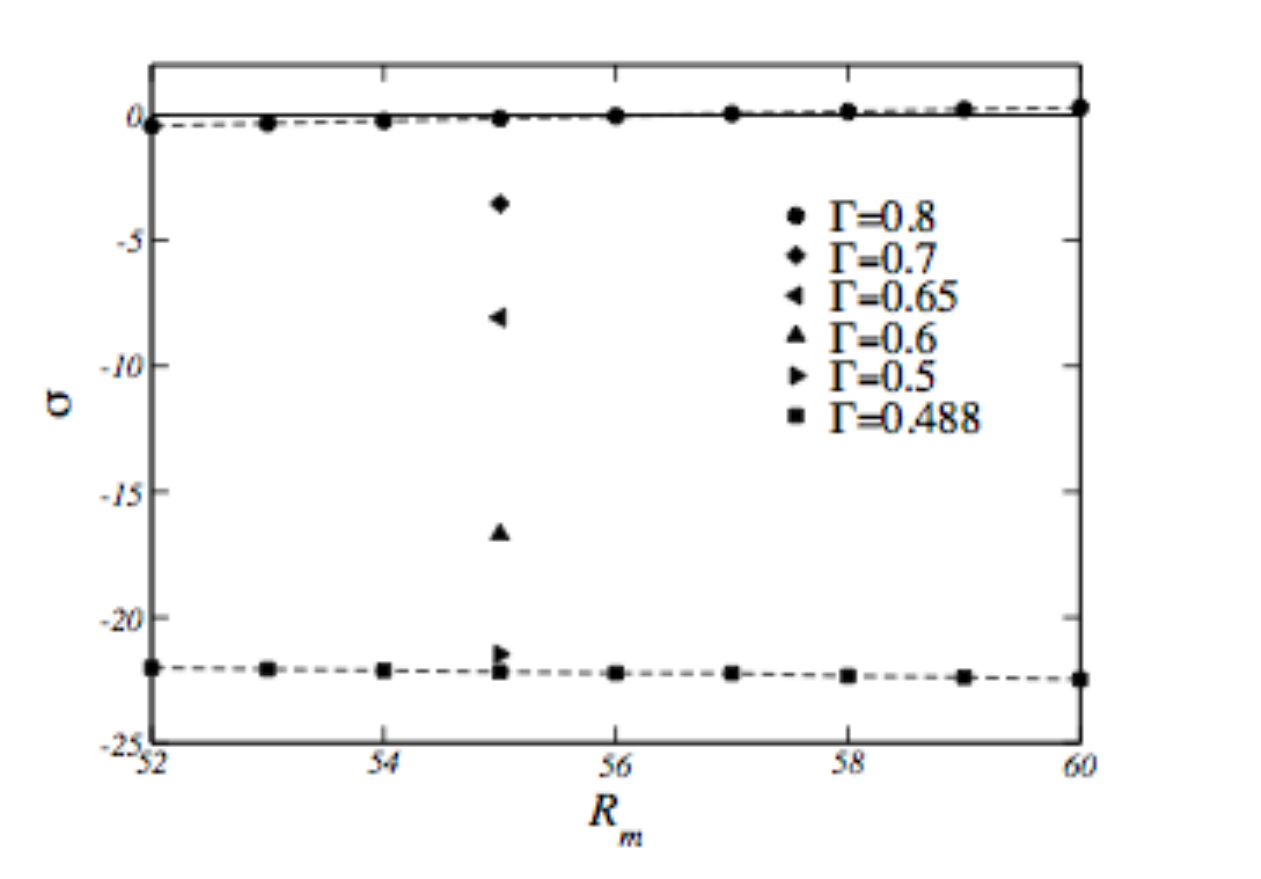}
\caption{Influence of the poloidal to toroidal ratio $\Gamma$ on the magnetic growth rate 
$\sigma$. {$\Gamma=0.488$} is the experimental value measured by SPIV in the experiment. 
The other cases correspond to artificial fields. A setup without annulus and (-) rotation
has been used. For the spline extrapolation we fixed nrp=4.}
\label{FIG13}
\end{figure}

These differences can be traced back to a difference in the poloidal to toroidal ratio $\Gamma$, 
that is much lower in the (-) case ($0.62$ with 
annulus and $0.49$ without annulus) than in the (+) case ($0.8$ with annulus or without annulus). This is in
agreement with previous studies showing that the optimum for dynamo growth in von Karman type flows is around 
$\Gamma \approx 0.75$ \cite{RCD05}. Fig. \ref{FIG13} exhibits the influence of $\Gamma$ on the magnetic growth rate 
$\sigma$ in the (-) case without annulus: the growth rate of the synthetic fields increases very sharply with $\Gamma$
between $\Gamma=0.5$ and $\Gamma =0.8$.

The insertion of the annulus exhibits two effects: (i) in the (+) case, it decreases slighly ($15 \%$) the critical 
magnetic Reynolds number while $\Gamma$ remains unchanged; (ii) in the (-) case, it increases the poloidal to toroidal ratio 
$\Gamma$ towards the range where dynamo action is available.

In the (+) case, both mean poloidal and toroidal flow increase when the annulus is inserted,
but their ratio $\Gamma$ does not change within the error bars. 
However, with the annulus, the axial gradient of the toroidal flow is larger 
in the region of the shear layer (cf. Fig. \ref{FIG7}): the gradient seems to be "less spread out" 
by the coherent structures. In fact, the annulus has been intruduced in the VKS experiment
to have an effect on the turbulent fluctuations, in particular
on the large scale fluctuations, by stabilizing the shear layer vortices in the annulus plane.
Indeed, the mean value of the normalized kinetic energy $\delta(t)$ decreases from $\bar{\delta} = 2.02$ to $1.48$ and 
its variance $\delta_2$ decreases from $0.18$ to $0.12$ (cf. \cite{CDM09}). This means that in the configuration with the annulus,
the instantaneous flow is much closer to the mean flow. This quantity has not been measured in the Riga and 
Karlsruhe experiments, but one can suspect that the instantaneous flow is very close to the mean flow, which explains
the very good agreement between kinematic dynamo simulations and experiments for these configurations. This is 
not the case for the VKS experiment, in which turbulent fluctuations are present at all scales. 
The annulus thus appears as a mean to reduce this problem, in particular for the large scales.

In the (-) case, $\Gamma$ has a low value without annulus, because both the toroidal part of the flow has increased 
and the poloidal part decreased. This low value is clearly out of the range for dynamo action in these flows. 
Moreover, the turbulent fluctuations are more intense in this sense of rotation and
the corresponding values of  $\bar{\delta}$ and $\delta_2$ are larger than in the (+) case. 
The insertion of the annulus has two effects.
First, it increases the poloidal to toroidal ratio, opening the possibility to get dynamo action, but still at large $R_m^c$. 
Second, it reduces $\bar{\delta}$ from $ 2.22$ to $1.50$ and its variance $\delta_2$ from $0.24$ to $0.11$ (cf. \cite{CDM09}), 
leading to values similar to the (+) case with annulus.
 
\begin{figure}
\includegraphics[clip=true,width=0.99\columnwidth]{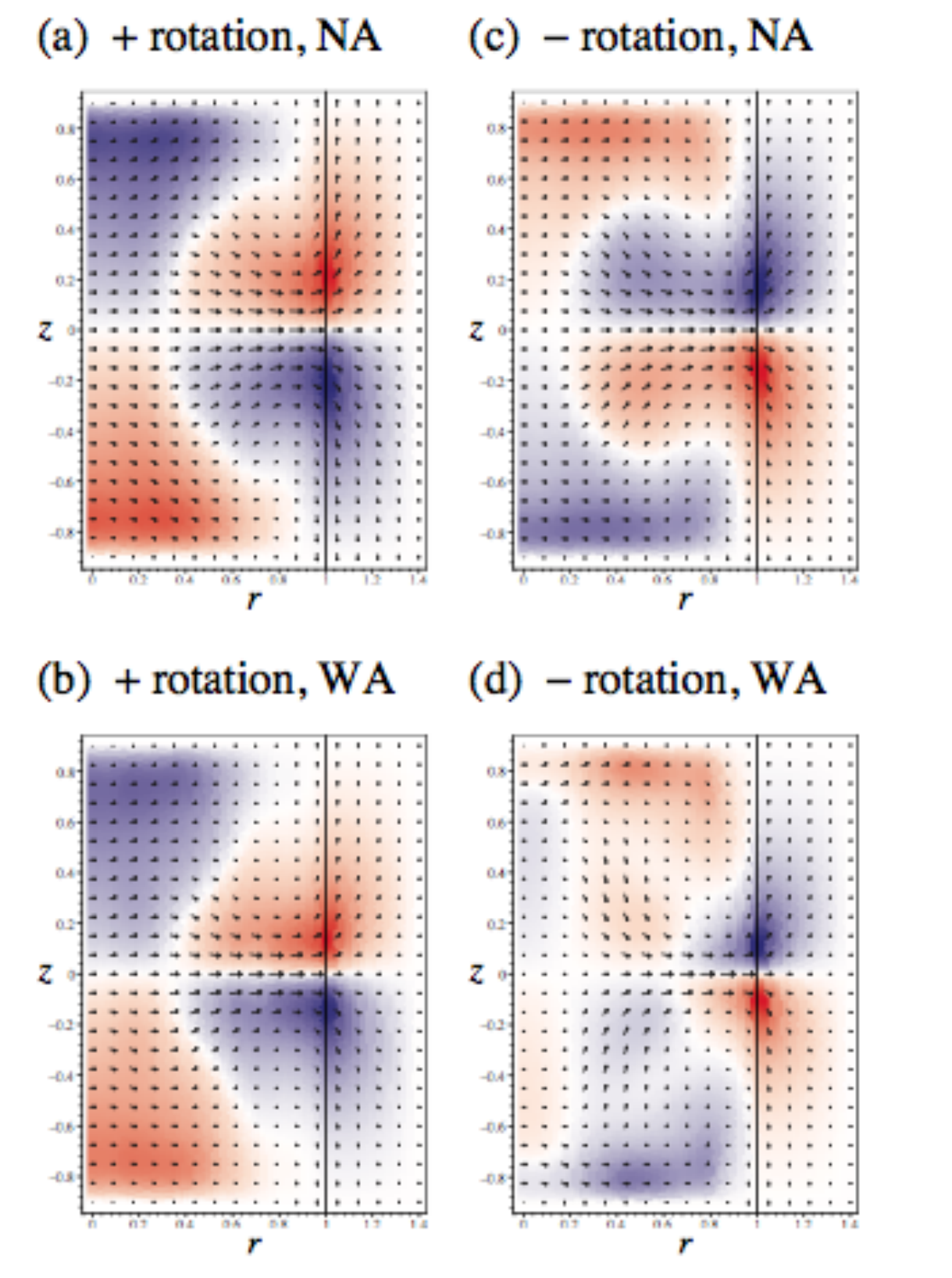}
\caption{Sections of the magnetic field $\mathbf{b}$ for different experimental setups, 
here for fixed $\Theta=0$. All configurations are with conducting layer 
$(w=0.4)$, whereas WA (resp. NA) specifies setups with (resp. without) annulus.
Arrows correspond to components lying in the cut plane and the color code to the component 
transverse to the cut plane, where red (resp. blue) denotes a positive (resp. negative) sign.
Parameters are (a) $R_m=44$, $\Gamma=0.8$, (b) $R_m=40$, $\Gamma=0.8$, (c) $R_m=60$, $\Gamma=0.8$ 
(artificial field), (d) $R_m=115$, $\Gamma=0.62$. For all cases nrp=4 has been used.}
\label{FIG14}
\end{figure}

\begin{figure}
\includegraphics[clip=true,width=0.99\columnwidth]{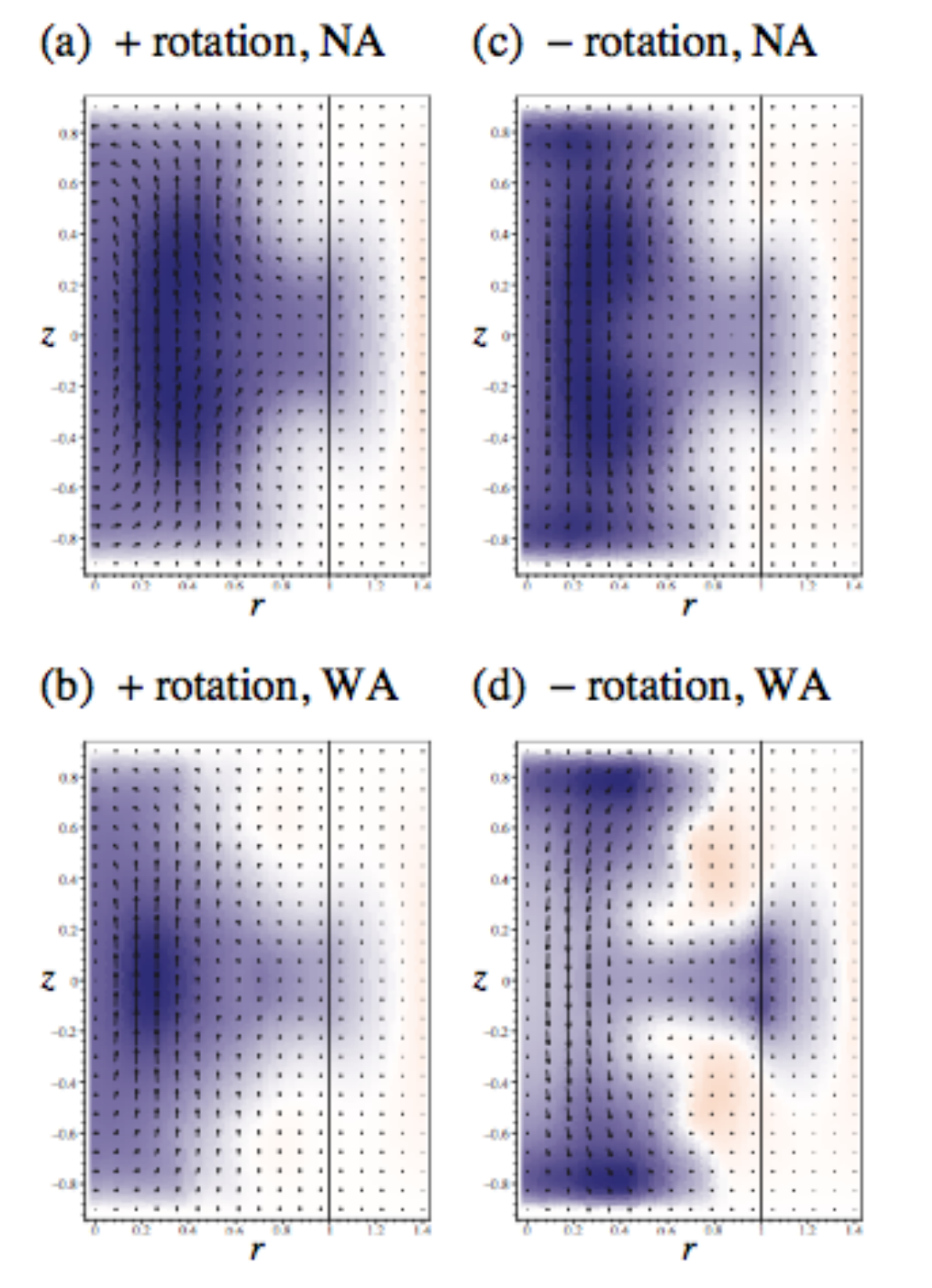}
\caption{Same as in Fig. \ref{FIG11}, for $\Theta=\pi/2$.}
\label{FIG15}
\end{figure}

{\it Magnetic growing mode.---}
The fastest growing mode we observe is always a $m=1$ mode. Recall that since our configuration and flow are
axisymmetric by construction and there are no external field source, the generation of a $m=0$ mode is
forbidden by Cowling's theorem \cite{M78}. 
Sections of the magnetic field obtained in the different configurations are presented in Fig. \ref{FIG14} and 
Fig. \ref{FIG15}. The configuration corresponding to (+) case without annulus is in agreement with the results of
\cite{RCD05}. The magnetic field in the (+) case with annulus reveals a sharper field....XXXXXXXX 

In the (-) case without annulus, the synthetic field with $\Gamma = 0.8$ looks very similar to the (+) case. 
On the contrary, the (-) case with annulus, that corresponds to a larger $R_m^c$ is very different from the 3 other cases.. 
XXXXXXXXXXXX

\subsection{Comparison with experiments}

{\it Onset of dynamo action.---}
When comparing these kinematic simulations results with experimental one, one observes that if the exact values 
are note recovered, the order of magnitude of the dynamo threshold $R_m^c$ as well as the general trend are correctly reproduced: 
with annulus, $R_m^c$ is larger for the (-) case than for the (+) case, and there is no dynamo in the (-) 
case without annulus in the available experimental $R_m$ range. 

The fact that the thresholds obtained in simulations are different from the experimental ones is not surprising:
the simulations presented in this paper have been performed with real mean flows but in an over-simplified VKS configuration. 
First of all, the lateral boundary conditions (shell of sodium at rest) have been taken into account but not the end boundaries 
(sodium behind the disks). However, the effect is in general to increase $R_m^c$ (\cite{SXG06}), whereas the 
experimental values are smaller than the numerical ones. 
Then, the effect of the soft-iron impellers (magnetic permeability $\sim 100$) has also been discarded. 
Using the synthetic MND flow \cite{MND06} to describe the von Karman mean axisymmetric flow, 
it has been shown that infinite permability boundary conditions are able to reduce the dynamo threshold  
by $5\%$ to $10\%$ \cite{GIFD08}. 
Finally, and perhaps more importantly, turbulence is not adressed in the simulations, whereas the Reynolds
number of the VKS experiment is between $10^6$ and $10^7$ \cite{P5}. Only the mean axisymmetric flow is 
used in the simulations : the large scale vortices of the shear layer as well as the blades vortices are not
considered, neither small scale turbulence. Non axisymmetric fluctuations, such as those generated by the flow
ejected by the blades of the impellers, could e.g. generate an $\alpha$ effect and give rise to an alpha-omega
type dynamo \cite{GAFDFauve}. The addition in the induction equation of an empirical alpha term localized 
in the vicinity of the impellers displays an axial dipole but the value of $\alpha$ required to obtain the 
experimental threshold appears to be unrealistic \cite{LNRLGP08}.

Very recent kinematic simulations with synthetic flows have been reported in two directions. 
First, by adding a modelization of the non axisymmetric vortices ejected by the  blades of the propellers 
to the synthetic MND mean axisymmetric flow, \cite{G09} has evidenced dynamo action with an axial 
dipole or quadrupole for $R_m^c \simeq 34$.  
Second, by adding a localized permeability distribution that mimics the shape of the impellers disk and blades,
\cite{GSG09} has found an equatorial dipole for $R_m^c \simeq 60$ without $\alpha$ effect and an axial dipole
for $R_m^c \simeq 32$ when including an $\alpha$ effect with a realistic value.

{\it Magnetic growing mode.---}
In all the studied cases the structure of the kinematic neutral mode is a $m=1$ equatorial dipole mode.
In the VKS experiment, the structure of the saturated has been reported in the (+) 
rotation sense, with annulus, and found to correspond mainly to a $m=0$ mode \cite{MBB07,P5}.
It would be interesting to have detailed measurements of the saturated mode structure in the VKS experiment, 
for the different configurations and in particular in the reverse rotation direction. One can e.g. wonder 
whether or not a $m=1$ component is present besides the axial dipole.

As reported above, recent simulations have exhibited an axial dipole (or quadrupole) with synthetic flows and
particular boundary conditions, and it would be of great interest to study the effect of the annulus and the 
direction of rotation on these results.

\section{Concluding remarks}

Through kinematic simulations using SPIV velocity fields measured in a water model experiment, we have shown 
that the rotation direction of the non axisymmetric impellers in the VKS experiment   
has a great impact on the threshold of the dynamo generated by the mean flow only.
By changing the rotation direction, one can increase the dynamo threshold by  a factor 3 in the case 
with annulus (1.4 in the VKS experiment), and a factor larger than 10 in the case without annulus
(larger than 1.6 in the VKS experiment). Even if the exact experimental results 
are not reproduced, the variation of the dynamo threshold with the direction sense is qualitatively reproduced 
by the kinematic simulations.

Since our kinematic simulations do not take into account the magnetic permeability due to iron disks-an effect 
that can decrease the threshold by a factor 2 (cf. \cite{GIFD08}) or generate a m=0 mode with an $\alpha$ effect
\cite{GSG09}, we do not expect our results to be more than qualitative. 
Furthermore, our linear simulations do not take into account the back reaction of the Lorentz force, 
that can enforce emergence of the m=0 mode as a secondary bifurcation of the original m=1 mode \cite{GDF08}. 
Finally, non axisymmetric turbulent fluctuations can play an important role in the generation of a magnetic 
dipole \cite{GAFDFauve,G09}.

Nevertheless, we have shown that the increase (or the absence) of the dynamo threshold observed in given 
configurations of the VKS experiment can have two origins: 
(i) the poloidal to toroidal ratio $\Gamma$, a parameter previously identified as crucial for dynamo action \cite{RCD05}. 
(ii) the normalized kinetic energy $\delta$ which quantifies the distance of the instaneous flow to the mean flow.
The parameter $\Gamma$ is certainly important for the alpha-omega mechanism proposed in \cite{GAFDFauve}, giving
the relative importance of the $\alpha$ effect generated by the blades vortices (poloidal recirculation)
vs. the omega effect due to differential rotation (toroidal flow).

Detailed experimental study of the propellers vortices dynamics with the rotation direction would represent an important
step to validate the numerical predictions. Work is in progress in this direction.

\section*{Acknowledgement}

\mbox{A.~Pinter} has been supported by the {\em Deutsche
Forschungsgemeinschaft}. We thank J.~L\'eorat for developing the numerical code, the VKE team F.~Ravelet, R.~Monchaux, P.~Diribarne
and P.-P.~Cortet for providing the experimental data and A.~Chiffaudel and C.~Normand 
for fruitful discussions.


\newpage

\begin{table}[h]
\begin{tabular}{|c|c|c|c||c|||c|c|}
\hline
\hline
Rotation sense & annulus & $\Gamma$ & $\bar{\delta}$ & $\delta_2$ & $R_m^c{\rm exp}$ & $R_m^c {\rm sim}$\\
\hline
\hline
+ & with    & $0.8$ & $1.48$ & $0.12$ & $\approx 32$ & $35.6-37.7$ \\
\hline
+ & without & $0.8$ & $2.02$ & $0.18$ & $\approx 32$ & $41.6-47.6$ \\
\hline
\hline
- & with & $0.62$ &  $1.50$ & $0.11$ & $\approx 43$ & $\approx 113$\\
\hline
- & without & $0.49$ & $2.22$ & $0.24$ & $>50$ & $>500$\\
\hline
- & without & $0.8$ & (*) & (*)   & (*) & $\approx 57$\\

\hline
\hline
\end{tabular}
\caption{Comparison of experimental and numerical critical magnetic Reynolds numbers for the different
VKS setups.  An interval for $R_m^c{sim}$ is given when different nrp have been used (cf. Fig. \ref{FIG9} 
and \ref{FIG11}), otherwise nrp=4. (*) denotes an artificial numerical field with a
prepared $\Gamma$ not available in experiments. The values of the mean normalized kinetic energy $\bar{\delta}$ 
and its variance $\delta_2$ are from \cite{CDM09}}
\label{TAB1}
\end{table}

\clearpage


\end{document}